\documentclass[a4paper,11pt]{article}
\pdfoutput=1

\usepackage{jcappub}

\usepackage[T1]{fontenc}

\usepackage{color}
\usepackage{graphicx}
\usepackage{dcolumn}
\usepackage{bm}
\usepackage{hyperref}
\usepackage{amsmath}
\usepackage{parskip}
\usepackage{adjustbox}
\hypersetup{colorlinks=true, linkcolor=magenta, filecolor=cyan, urlcolor=blue}

\def\@Aboxed#1&#2\ENDDNE{%
  \settowidth\@tempdima{$\displaystyle#1{}$}%
  \addtolength\@tempdima{\fboxsep}%
  \addtolength\@tempdima{\fboxrule}%
  \global\@tempdima=\@tempdima
  \kern\@tempdima
  &
  \kern-\@tempdima
  \fcolorbox{red}{yellow}{$\displaystyle #1#2$}
}

\newlength\dlf

\title{The Ashtekar Variables and a Varying Cosmological Constant from Dynamical Chern-Simons Gravity}

\author[a]{Stephon Alexander,}
\author[a]{Tatsuya Daniel,}
\author[b]{and João Magueijo}

\affiliation[a]{Brown Theoretical Physics Center, Department of Physics, Brown University,\\ Providence, RI 02912, USA}
\affiliation[b]{Theoretical Physics Group, The Blackett Laboratory, Imperial College,\\Prince Consort Rd., London, SW7 2BZ, United Kingdom}

\emailAdd{stephon\_alexander@brown.edu}
\emailAdd{tatsuya\_daniel@brown.edu}
\emailAdd{j.magueijo@imperial.ac.uk}

\abstract{We revisit the Kodama state by quantizing the theory of General Relativity (GR) with dynamical Chern-Simons (dCS) gravity. We find a new exact solution to the Wheeler-DeWitt equation where the Pontryagin term induces a modification in the Kodama state from quantizing GR alone. The dCS modification directly encodes the variation of the cosmological constant $\Lambda$. }

\begin{document}
\maketitle
\flushbottom

\section{\label{sec:background}Introduction}

The pioneering work of Ashtekar opened a new possibility of expressing GR in terms of a complex self-dual SU(2) connection $A_a^i$, and its conjugate momentum $E_a^i$ \cite{PhysRevLett.57.2244}. From these so-called Ashtekar variables, Kodama \cite{Kodama90} then found an exact solution to a non-perturbative quantization of GR with a positive cosmological constant, called the \emph{Kodama state}. The Kodama state is obtained by solving the Wheeler-DeWitt equation, which describes the quantum version of the Hamiltonian constraint, a linear combination of spatial and time diffeomorphism constraints that reflects the reparametrizability of GR under spatial and time coordinates. Solving the Wheeler-DeWitt equation amounts to finding the ground state of the theory; while it has proven to be insurmountable - with exception to minisuperspace approximations - it is tractible using Ashtekar variables.

Ashtekar's formalism thus opened a new approach to quantizing gravity non-perturbatively; other approaches include loop quantum gravity, causal dynamical triangulations, spin foam and causal sets, which are non-perturbative approaches that attempt to yield a self-consistent quantization of gravity. Many of these approaches have had difficulties in making contact to a semi-classical limit. Moreover, while the Kodama state is a candidate for describing a ground state of quantum gravity with a positive cosmological constant non-perturbatively, it suffers from a number of issues, including non-normalizability, its CPT-violating properties (and consequent impossibility of having a positive energy), and lack of gauge invariance under large gauge transformations \cite{Magueijo_2021}.

Nevertheless, the Ashtekar variables have faciliated formulating non-perturbative quantizations in extensions of GR as well. There are well-motivated reasons to consider modifications to GR, from anomaly cancellation to explaining observations such as leptogenesis and parity violation. Recent work has parametrized gravitational wave propagation in extensions of GR, which allows one to map theory parameters to observables \cite{daniel2024}.  In particular, dynamical Chern-Simons gravity (dCS) is a strong candidate to provide a modification to GR, a theory which has been studied and developed extensively \cite{Alexander_2009}. It has been theorized that chiral gravitational waves could emerge from dCS, where a change in a left-handed wave sources the right-handed wave and vice versa. Chiral gravitational waves could be a way to resolve the argument that the Kodama state is unphysical due to negative helicity states in the expansion of the Kodama state that result in negative energies; a state with negative energy could be recast as a positive-energy state with opposite helicity \cite{https://doi.org/10.48550/arxiv.gr-qc/0306083,Freidel_2004}. 

In this note, we revisit the Kodama state by adding dCS into the gravitational theory and therefore into the Wheeler-DeWitt equation. We find a new exact solution to the Hamiltonian constraint such that the dCS modification allows for the time variation of the cosmological constant $\Lambda$. 

\section{The New Hamiltonian Constraint and Wavefunction}
We begin by considering the action Eq.~(12) of \cite{https://doi.org/10.48550/arxiv.1807.01381}, that of GR coupled to $N$ chiral fermion fields, all expressed in terms of Ashtekar variables (here $-\frac{3}{16\pi\Lambda G} \equiv \theta$) \cite{samuel87, jacobsonsmolin87}:
\begin{align}
    S = \int_M\frac{1}{8\pi G}\bigg\{\epsilon^{abcd}e_a \wedge e_b \wedge e_c \wedge e_d \bigg(R + 2\Lambda\bigg) + \theta R^{AB} \wedge R_{AB}\bigg\} + \Sigma_N\overline{\Psi}_{A'}\sigma_a^{A'B}e^a \wedge (\mathcal{D}\Psi)_B, \label{eq:1}
\end{align}
where we fix a topology $M = I \times \Sigma$, $I$ being the interval and we take $\Sigma$ to be compact. Notice the addition of the $\theta R^{AB} \wedge R_{AB}$ term in Eq.~(\ref{eq:1}).
\newline
Applying a 3+1 decomposition to the action yields \cite{https://doi.org/10.48550/arxiv.1807.01381} 
\begin{align}
    S = \int dt\bigg[\int_{\Sigma}\frac{1}{8\pi G}\bigg(E^{ai}\dot{A}_{ai} - \mathcal{N}\mathcal{H} - \mathcal{N}^i\mathcal{D}_a - \mu_i\mathcal{G}^i\bigg) + \frac{3}{16\pi G}\frac{\dot{\Lambda}}{\Lambda^2}\int_{\Sigma}Y_{\text{CS}}(A)\bigg], \label{eq:action}
\end{align}
where $\mathcal{N}$ is the lapse function, $\mathcal{H}$ is the Hamiltonian constraint that generates the rest of the group of the spacetime (and thus changes of the slicings of spacetime into spatial slices), $\mathcal{N}^a$ is the shift function, $\mathcal{D}_a$ generates the diffeomorphisms of $\Sigma$, $\mathcal{G}^i$ generates the local gauge transformations, and $Y_{\text{CS}}(A) = \epsilon^{abc}F_{bc}^iA_{ai}$ is the Chern-Simons invariant of the Ashtekar connection. We are taking $\theta$ (equivalently $\Lambda$) to be a function of time only. 
\newline 
\indent 
The last term in Eq.~(\ref{eq:action}) is obtained by writing the Riemann tensor $R_i^j = d\omega_i^j + \omega_i^k \wedge \omega_k^j$ (where $\omega$ is the four-dimensional spin connection) in terms of the complex, self-dual SU(2) connection $A$ using the fact that\footnote{In other words, $A$ is the three-dimensional spatial projection of $\omega$.}
\begin{equation}
    A_a^i \equiv -\frac{1}{2}\epsilon^{ij}_{~~k}\omega_{aj}^k - i\omega_{a0}^{~~i}.
\end{equation}
Given that the Hamiltonian constraint must be gauge invariant and a scalar (since the parameter it multiplies is proportional to the local change in the time coordinate), the two simplest terms it can contain, which are lowest order in derivatives, are
\begin{equation}
    \epsilon_{abc}\text{Tr}E^aE^bE^c~~~~~\text{and}~~~~~\text{Tr}E^aE^bF_{ab}, \label{eq:trace}
\end{equation}
where the trace is in some lie algebra $G$. If we take the simplest nontrivial choice for $G$, which is $SU(2)$, the two terms in Eq.~(\ref{eq:trace}) give Einstein's equations.
\newline
\indent
Therefore, for the Hamiltonian corresponding to GR we write \cite{smolin02},
\begin{align}
    \mathcal{H_{\text{GR}}} = \frac{1}{8\pi G}\epsilon_{ijk}E^{ai}E^{bj}\bigg(F_{ab}^k + \frac{\Lambda}{3}\epsilon_{abc}E^{ck}\bigg), \label{eq:3}
\end{align}
and we add to it a Pontryagin term $\theta R^{AB} \wedge R_{AB}$. For this we can simply read off of Eq.~(\ref{eq:1}); from \cite{https://doi.org/10.48550/arxiv.1807.01381}, we see that the Pontryagin term contributes an additional term $\frac{3}{16\pi G}\frac{\dot{\Lambda}}{\Lambda^2}Y_{CS}(A)$ to the GR Hamiltonian. Using the fact that $\theta = -\frac{3}{16\Lambda\pi G}$, we have
\begin{align}
    \frac{\dot{\Lambda}}{\Lambda^2} &= \frac{16\pi G}{3}\dot{\theta}.
\end{align}
Thus, the extra term in the Hamiltonian is 
\begin{align}
    \mathcal{H}_{\text{DCS}} = \frac{3}{16\pi G}\frac{\dot{\Lambda}}{\Lambda^2}Y_{CS}(A) = \dot{\theta}Y_{CS}(A). \label{eq:5}
\end{align}
We apply the quantization procedure
\begin{align}
    \hat{E}^{ai} \rightarrow - G\frac{\delta}{\delta A_{ai}},
\end{align}
and the Hamiltonian constraint becomes
\begin{align}
    &(\mathcal{H}_{\text{GR}} + \mathcal{H}_{\text{DCS}})\Psi(A) = 0 \\
    \Rightarrow~-&\dot{\theta}Y_{\text{CS}}(A)\Psi(A) = \frac{G}{8\pi}\epsilon_{ijk}\frac{\delta}{\delta A_{ai}}\frac{\delta}{\delta A_{bj}}\bigg(F_{ab}^k - \frac{G\Lambda}{3}\epsilon_{abc}\frac{\delta}{\delta A_{ck}}\bigg)\Psi(A). \label{eq:8}
\end{align}
For $\Psi(A)$ we try the ansatz
\begin{align}
    \Psi(A) = \psi_K(A)\chi(A,\dot{\theta}), \label{eq:9}
\end{align}
where $\psi_K(A) \propto \text{exp}\bigg(\frac{3}{2G\Lambda}\int_{\Sigma} Y_{\text{CS}}(A)\bigg)$ is the solution to the Wheeler-DeWitt equation when the left-hand side of Eq.~(\ref{eq:8}) is zero (i.e.~without the inclusion of the Pontryagin term).
\newline
Plugging in our ansatz Eq.~(\ref{eq:9}) into Eq.~(\ref{eq:8}), we have
\begin{align}
    -\dot{\theta}Y_{\text{CS}}\psi_K\chi = \epsilon_{ijk}\frac{G}{8\pi}\frac{\delta}{\delta A_{ai}}\frac{\delta}{\delta A_{bj}}\bigg(F_{ab}^k - \frac{G\Lambda}{3}\epsilon_{abc}\frac{\delta}{\delta A_{ck}}\bigg)\psi_K\chi. \label{eq:10}
\end{align}
We make the assumption that we are working with non-degenerate metrics, that is, $E \cdot E \sim \text{det}~g$. We also note that the terms in parentheses on the right-hand side of Eq.~(\ref{eq:10}) annihilate $\psi_K$ by definition, and moreover we know that $G = M_P^{-2}$, where $M_P$ is the Planck mass. This means that Eq.~(\ref{eq:10}) reduces to 
\begin{align}
    \frac{\Lambda}{24\pi M_P}\chi' + \bigg(\dot{\theta}Y_{CS}(A) + \frac{\epsilon_{ijk}M_PE^{ai}E^{bj}F_{ab}^k}{8\pi}\bigg)\chi = 0, \label{eq:14}
\end{align}
where the primes denote derivatives with respect to $A$.
\newline 
\indent
Here we also make the assumption that $F_{ab}^k$ is independent of $A$. This means we can set $F_{ab}^k$ equal to a constant; setting this constant equal to zero for simplicity, rearranging Eq.~(\ref{eq:14}) and using the standard technique of separation of variables (again assuming that $E \cdot E \sim \text{det}~g$), we find 
\begin{align}
    \chi(A,\dot{\theta}) = \text{exp}\bigg(-\frac{24\pi M_P\dot{\theta}}{\Lambda}\int_{\Sigma} Y_{\text{CS}}(A)\bigg). \label{eq:19}
\end{align}
Thus, the full wavefunction is
\begin{align}
    \Psi(A) &= \psi_K(A)\chi(A,\dot{\theta})  \\ &= \text{exp}\bigg[\frac{3M_P^2}{2\Lambda}\bigg(1 - \frac{16\pi\dot{\theta}}{M_P}\bigg)\int_{\Sigma} Y_{\text{CS}}(A)\bigg], \label{eq:23}
\end{align}
or in terms of $\dot{\Lambda}$,
\begin{align}
    \Psi(A) = \text{exp}\bigg[\frac{3M_P^2}{2\Lambda}\bigg(1 - \frac{3\dot{\Lambda}}{\Lambda^2}M_P\bigg)\int_{\Sigma} Y_{CS}(A)\bigg]. 
    \end{align}
\newline
\indent
Eq.~(\ref{eq:19}) is the modification to the Kodama state with the inclusion of the Pontryagin term (via Eq.~(\ref{eq:9})). 

\section{Gauss and Diffeomorphism Constraints}
We now want to see if the Gauss and spatial diffeomorphism constraints hold when the Kodama state gets modified with the Pontryagin term\footnote{The Gauss constraint is essentially Gauss's Law, which expresses local in-space rotational invariance, the origin of the SU(2) gauge invariance.}. From Eq.~(2.5) of \cite{Soo_2002}, the Gauss constraint is 
\begin{align}
    D_i^AE^{ia} = 0, \label{eq:gaussconstraint}
\end{align}
and the diffeomorphism constraint is
\begin{align}
    \epsilon_{ijk}E^{j}_aB^{ia} = 0, \label{eq:diffconstraint}
\end{align}
where \{$E^{ia}(\textbf{x}), A_{jb}(\textbf{y})\} = i(8\pi G)\delta^i_j\delta^a_b\delta^3(\textbf{x} - \textbf{y})$. 
\newline
We can define an operator $\hat{Q}^{ia}$ where
\begin{align}
    \hat{Q}^{ia} \equiv B^{ia} + \frac{\lambda}{3}\hat{E}^{ia}. \label{eq:Q}
\end{align}
Plugging Eq.~(\ref{eq:Q}) into Eq.~(\ref{eq:gaussconstraint}) yields, when acting on $\Psi(A)$,
\begin{align}
    D_i^A\hat{Q}^{ia}\Psi(A) = \frac{1}{2}D_i^A\epsilon^{ijk}F_{jk}^a\Psi(A), \label{eq:29}
\end{align}
where we used the fact that $B^{ia} = \frac{1}{2}\epsilon^{ijk}F_{jk}^a$. Since $D_i^A\epsilon^{ijk}F_{jk}^a = 0$, we have that
\begin{align}
    D_i^A\hat{Q}^{ia}\Psi(A) = 0. \label{eq:30-1}
\end{align}
As for the diffeomorphism constraint, we use Eq.~(\ref{eq:Q}) again, this time in Eq.~(\ref{eq:diffconstraint}), to find 
\begin{align}
    \epsilon_{ijk}\hat{E}_a^j\hat{Q}^{ia}\Psi(A) &= \frac{\lambda}{3}\epsilon_{ijk}\text{Tr}(\hat{E}^j\hat{E}^i)\Psi(A),
\end{align}
which vanishes due to the symmetry of the frame fields. Therefore, we have
\begin{align}
    \epsilon_{ijk}\hat{E}_a^j\hat{Q}^{ia}\Psi(A) = 0. \label{eq:34}
\end{align}
Eqs.~(\ref{eq:30-1}) and (\ref{eq:34}) are exactly Eqs.~(4.3) of \cite{Soo_2002}. The Kodama state $\psi_K(A)$ satisfies Eqs.~(\ref{eq:30-1}) and (\ref{eq:34}), by definition. However, since these constraints are trivially satisfied, this means that the modified Kodama state Eq.~(\ref{eq:23}) would solve these constraints without modifying them. This also means that any $\Psi$ (and thus any modification to the Kodama state from GR) would satisfy the Gauss and diffeomorphism constraints unmodified, but $\Psi$ would have to solve the new modified Hamiltonian constraint (Eq.~(\ref{eq:8}) in this case).
\newline
\indent
In our modified theory, one can investigate the symplectic structure of the algebra of constraints and whether or not it has closure. We leave this analysis for future work.

\section{The Full Solution}
Now we consider solving Eq.~(\ref{eq:14}) without making the assumption that $F$ is independent of $A$. We have
\begin{align}
    \frac{1}{\chi}\chi' = -\frac{24\pi M_P}{\Lambda}\bigg(\dot{\theta}Y_{CS}(A) + \frac{\epsilon_{ijk}M_PE^{ai}E^{bj}F_{ab}^k}{8\pi}\bigg),
\end{align}
so
\begin{align}
    \chi(A,\dot{\theta}) = \text{exp}\bigg[&-\frac{24\pi M_P\dot{\theta}}{\Lambda}\int_{\Sigma} Y_{CS}(A) - \frac{3M_P^2}{\Lambda}\int_{\Sigma}\epsilon_{ijk}E^{ai}E^{bj}F_{ab}^k\bigg], \label{eq:24}
\end{align}
and the full Kodama state is
\begin{align}
    \Psi(A) &= \psi_K(A)\chi(A,\dot{\theta}) \\
    &= \text{exp}\bigg\{\frac{3M_P^2}{\Lambda}\bigg[\frac{1}{2}\bigg(1 - \frac{16\pi\dot{\theta}}{M_P}\bigg)\bigg(\int_{\Sigma}Y_{CS}(A)\bigg) - \bigg(\int_{\Sigma}\epsilon_{ijk}E^{ai}E^{bj}F_{ab}^k\bigg)\bigg]\bigg\}. \label{eq:27}
\end{align}

\begin{figure}
    \centering
    \includegraphics[width=0.7\textwidth]{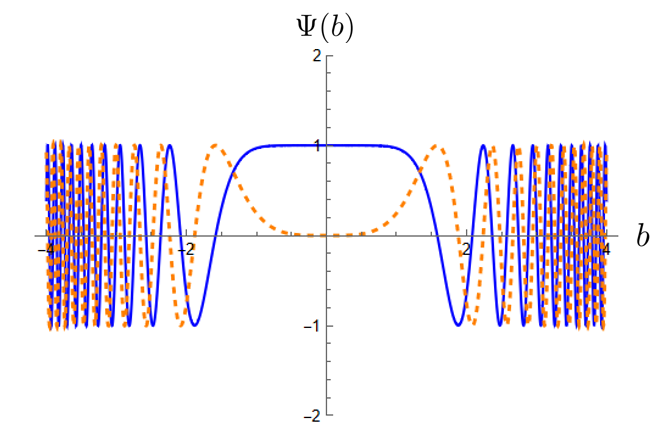}
    \caption{The real and imaginary parts of the minisuperspace wavefunction Eq.~(\ref{eq:30}) (solid blue and dashed orange, respectively) for $c=0$.}
    \label{fig:kodama}
\end{figure}

We can reduce to minisuperspace by making the general ansatz consistent with homogeneity and isotropy:
\begin{align}
    A_a^i &= \delta_a^i(ib+c), \label{eq:28} \\
    E_i^a &= \delta_i^aa^2, \label{eq:29}
\end{align}
where $b$ and $c$ are functions of time. Using Eqs.~(\ref{eq:28}) and (\ref{eq:29}) in (\ref{eq:27}) yields the minisuperspace wavefunction
\begin{align}
    \Psi_b(b) = \mathcal{N}\text{exp}\bigg\{-\frac{9M_P^2}{\Lambda}\bigg[\bigg(1 - \frac{16\pi\dot{\theta}}{M_P}\bigg)\bigg\{bc^3 + \frac{3}{2}ib^2c^2 - b^3c - \frac{i}{4}b^4\bigg\} + a^4\bigg\{bc^2 + ib^2c - \frac{b^3}{3}\bigg\}\bigg]\bigg\}, \label{eq:30}
\end{align}
where $\mathcal{N}$ is the normalization constant.
Reducing Eq.~(\ref{eq:action}) to minisuperspace and identifying the commutation relation
\begin{equation}
    [\hat{b},\hat{a}^2] = \frac{i}{3V_cM_P^2},
\end{equation}
we have
\begin{align}
    \Psi_{a^2}(a^2) &= \mathcal{N}M_P\sqrt{3V_c}\int \frac{\text{d}b}{\sqrt{2\pi}}e^{-3iV_cM_P^2a^2b}\Psi_b(b). \label{eq:32}
\end{align}
If we set $c = 0$, expand the exponential in Eq.~(\ref{eq:30}) using the power series expansion $e^{x} = \sum_{n=0}^{\infty} \frac{x^n}{n!}$, and take the integral over the real number line, Eq.~(\ref{eq:32}) becomes
\begin{align}
   \Psi_{a^2}(a^2) = \frac{\mathcal{N}'}{(a^2)^2}\bigg[\bigg(1 - \frac{16\pi\dot{\theta}}{M_P}\bigg)\frac{1}{V_cM_P^2(a^2)^3} - 1\bigg] + \ldots, \label{eq:33}
\end{align}
where 
\begin{align}
    \mathcal{N}' = -\frac{2\mathcal{N}}{9\Lambda V_c^4M_P^5}\sqrt{\frac{3V_c}{2\pi}}.
\end{align}
When $c = 0$, the terms at each order in the expansion in Eq.~(\ref{eq:33}) are much smaller than the previous order if $|\frac{9M_P^2}{\Lambda}| \ll 1$, i.e. if $\Lambda \ll M_P^2$. The minisuperspace wavefunction Eq.~(\ref{eq:30}) is plotted in Fig.~\ref{fig:kodama} for $c=0$.
\newline
\indent
It is difficult to make contact with the minisuperspace wavefunction derived in~\cite{Magueijo_2020} for the Pontryagin theory, because the wavefunction derived there always contains an admixture of Gauss-Bonnet contributions. 

\section{Discussion}
In quantizing the theories of GR and dCS with a cosmological constant $\Lambda$, the Pontryagin term that arises from dCS ($\theta R\Tilde{R}) $ yields a modification to the Kodama state from quantizing GR alone. The variation of the Chern-Simons coupling $\dot{\theta}$ can be directly related to $\dot{\Lambda}$. The full exact solution depends on the value of the three-curvature integrated over the connection space and the value of the Chern-Simons functional $Y_{\text{CS}}(A)$ integrated over the same space. 
\newline
\indent
We can define a Chern-Simons time $\tau_{\text{CS}} = \int Y_{\text{CS}}(A)$ as in \cite{https://doi.org/10.48550/arxiv.1807.01381,notime}, where $\Lambda$ and $\tau_{\text{CS}}$ obey the commutation relations
\begin{align}
    [\hat{\Lambda}, \hat{\tau}_{\text{CS}}] = i\frac{16\pi\Lambda^2}{3M_P^2}, \label{eq:35}
\end{align}
or replacing $\Lambda$ with $\theta$, 
\begin{align}
    \bigg[\hat{\theta}, \hat{\tau}_{\text{CS}}\bigg] = i.
\end{align}
Furthermore, an uncertainty principle between $\Lambda$ and $\tau_{\text{CS}}$ can be derived:
\begin{align}
    \Delta\Lambda\Delta\tau_{\text{CS}} \geq \frac{4\pi}{3M_P^2}\langle \hat{\Lambda} \rangle^2. \label{eq:46}
\end{align}
From Eq.~(\ref{eq:46}), we see that $\Lambda$ and $\tau_{\text{CS}}$ are incompatible with each other if $\langle \hat{\Lambda} \rangle$ is large (correspondingly, if $\langle\hat{\theta}\rangle$ is small). On the other hand, the cosmological constant and Chern-Simons time can be treated as classical variables if $\langle \hat{\Lambda} \rangle$ is small (i.e. $\langle\hat{\theta}\rangle$ is large) \cite{https://doi.org/10.48550/arxiv.1807.01381}. This can be achieved by tuning the Chern-Simons coupling $\theta$ such that it is sufficiently small. 
\newline
\indent

\section{Summary and Outlook} 
We have found that the correction from dCS gravity directly allows for the variation of $\Lambda$, which makes our full modified Kodama state a strong candidate for describing quantum gravity non-perturbatively with a time-varying cosmological constant.
\newline
\indent
We close by noting that other classical methods for inducing a time variation in $\Lambda$ have been proposed, not always phenomenologically successful~\cite{lee1,lee2,zlos1,zlos2}. This early work, nonetheless, already points to a deep relation between a varying $\Lambda$ and the Pontryagin term, as well as Chern-Simons gravity. The relation with their metric representation cousins of these states~\cite{CSHHV}, however, remains to be explored. One senses that the decades-long discussion about boundary conditions initiated by Hartle and Hawking~\cite{HH} and Vilenkin~\cite{Vil} could benefit from this radically different perspective.

\acknowledgments
We would like to thank Gabriel Herczeg and Tucker Manton for helpful comments related to this paper. This work was supported by the Simons Foundation Mathematics and Physical Sciences (MPS) division, and by the STFC Consolidated Grant ST/T000791/1.

\bibliographystyle{unsrtnat}
\bibliography{dCS_Kodama_State}

\end{document}